%% file: main.tex
\documentclass{article}
\usepackage{spconf,amsmath,graphicx}

\usepackage{multirow}
\usepackage[nolist]{acronym}
\usepackage{amsfonts}
\usepackage{xcolor}
\usepackage{amsmath}
\usepackage{graphicx}
\usepackage[normalem]{ulem}
\usepackage{footnote} 
\usepackage{subfigure} 
\usepackage{url}
\usepackage{multirow}
\usepackage{amssymb}
\usepackage{tabularx} 

\usepackage[normalem]{ulem}
\useunder{\uline}{\ul}{}

\newcommand{\eg}{\text{e.g. }}

\newcommand{\ie}{\text{i.e. }}

\newcommand{\fig}[1]{Fig.~}
\newcommand{\tab}[1]{Tab.~}
\newcommand{\Sec}[1]{Sec.~}
\newcommand{\eq}[1]{Eq.~}

\begin{document}
\include{acro}
\ninept

\title{Multi-target DoA estimation with an audio-visual fusion mechanism}

\name{Xinyuan Qian, Maulik Madhavi, Zexu Pan, Jiadong Wang, Haizhou Li\thanks{This research work is  supported by the Neuromorphic Computing project, Programmatic Grant No. A1687b0033 from the Singapore Government’s Research, Innovation and Enterprise 2020 plan (Advanced Manufacturing and Engineering domain), and Human-Robot Interaction Phase 1 (Grant No. 192 25 00054) from the National Research Foundation, Prime Minister’s Office, Singapore under the National Robotics Programme.}}
\address{Department of Electrical and Computer Engineering,\\
  National University of Singapore, Singapore}
  




\maketitle

\begin{abstract}
Most of the prior studies in the spatial \ac{DoA} domain focus on a single modality.
However, humans use auditory and visual senses to detect the presence of sound sources. With this motivation, we propose to use neural networks with audio and visual signals for multi-speaker localization. 
The use of heterogeneous sensors can provide complementary information to overcome uni-modal challenges, such as noise, reverberation, illumination variations, and occlusions. We attempt to address these issues by introducing an adaptive weighting mechanism for audio-visual fusion. We also propose a novel video simulation method that generates visual features from noisy target 3D annotations that are synchronized with acoustic features. 
Experimental results confirm that audio-visual fusion consistently improves the performance of speaker DoA estimation, while the adaptive weighting mechanism shows clear benefits.

\end{abstract}



\section{Introduction}
In human-robot interaction, a robot relies on its \ac{SSL} mechanism to direct its attention. 
Traditionally, \ac{SSL} approaches only use audio signals
and attempt as a signal processing problem 
\cite{knapp1976generalized,brandstein1997robust,schmidt1986multiple}.
However, those approaches are adversely affected by acoustically challenged conditions,
such as noise and reverberation scenarios \cite{he2018deep}. To address that, several \ac{NN}-based approaches were explored \cite{chakrabarty2019multi,adavanne2018direction,he2018deep,pan2020multitones} assuming a sufficient amount of data are available.
Specifically,  location-related \ac{STFT} cues are mapped to sound \ac{DoA} information in \cite{chakrabarty2019multi,adavanne2018direction} while  the \ac{GCC-PHAT} cues are used in \cite{he2018deep,pan2020multitones}.
Despite the progress, many research problems remain. One of them is multi-speaker localization in real multi-party human-robot interaction scenarios under acoustic challenging conditions \cite{he2018deep}.

Considering seeing and hearing are the two most essential human cognitive abilities,  studies observed that audio and video convey complementary information and may help to overcome uni-modal limitations of a degradation condition for scene analysis \cite{katsaggelos2015audiovisual,atrey2010multimodal,shivappa2010audiovisual}.
There is a very broad literature of audio-visual approaches for speaker localization over the past decades \cite{beal2003a,qian2019multi,ban2019variational}. However, it was not until recently that the deep learning-based approaches have attracted more attention, thanks to the increasing computational power and rapid development in \ac{NN} techniques.
Nevertheless, most of these methods aim at locating sound sources in visual scenes \cite{senocak2018learning,tsiami2020stavis,tian2018audio, ramaswamy2020see}. Specifically, an attention mechanism is incorporated into the individual sound and vision network to model the audio-visual image correspondence \cite{senocak2018learning} . 
A visual saliency network is employed in \cite{tsiami2020stavis}, together with an audio representation network, to feature a \ac{SSL} module for producing an audio-visual saliency map. 
An attention network is proposed in \cite{tian2018audio} to  learn the visual regions of a sounding event.
By fusing audio and visual features using LSTM and bilinear pooling,  the audio assisted visual feature extraction is described in \cite{ramaswamy2020see}.  All the research studies use audio as a supplementary modality for visual localization and require the sound sources to be both audible and visible.

Unlike the prior studies, we aim to perform audio-visual speaker localization in the spatial \ac{DoA} domain where targets can appear either inside (visible) or outside (invisible) the camera's \ac{FoV}.
We  propose two neural network architectures and make the following contributions in this paper: (1) we propose a novel video simulation method to deal with the lack of video data; (2) 
for the first time, we design a deep learning network for audio-visual multi-speaker \ac{DoA} estimation, and (3) we adopt an adaptive weighting mechanism in a simple feedforward network to estimate the multi-modal reliability under different conditions. 

\section{Proposed Method}
\label{sec:proposed_method}


Given a sequence of frame-synchronized audio and video signals 
captured by a microphone array and a calibrated camera, we aim to estimate the \ac{DoA} information $\theta=[-180^\circ,180^\circ)$ for each sound source at each frame. Next, we describe the way we characterize audio and video signals, the video simulation method, and the proposed neural networks.

\subsection{Audio features}
The \ac{GCC-PHAT} is widely used to calculate the time different of arrival (TDOA) between any two microphones in a microphone array~\cite{he2018deep,pan2020multitones}. 
We adopt it as the audio feature ~\cite{knapp1976generalized} due to its robustness in the noisy and reverberant environment \cite{florencio2008does} and the fewer tunable parameters than the other counterparts \eg \ac{STFT} \cite{chakrabarty2019multi}.
Let $S_l$ and $S_p$ be the Fourier transforms of audio sequence at $l$ and $p^{th}$ channels of the microphone array, respectively. We compute the GCC-PHAT features with different delay lags $\tau$ as:
\begin{equation}\label{eq:gccphat}
    \text{GCC-PHAT}_{lp}(\tau) = \sum_{k}\mathcal{R}\left(\frac{S_l[k](S_p[k])^{*}}{|S_l[k](S_p[k])^{*}|} e^{j \frac{2\pi k}{N} \tau}\right)
\end{equation}
where $*$ denotes the complex conjugate operation, $\mathcal{R}$ denotes the real part of complex number and $N$ denotes the FFT length. Here, the delay lag $\tau$ between  two signals arrived is reflected in the steering vector  $e^{j \frac{2\pi k}{N} \tau}$ in \eq~\ref{eq:gccphat}. 
\vspace{-0.1cm}
\subsection{Visual features and simulation}\label{ssec:videofeature}
With the advent of deep learning, accurate face detection at low computational cost becomes widely available ~\cite{zou2019object}. Let us define ${\bf b}_{d}=(u,v,w,h)^\intercal_{d}$ as the face detection bounding box $d \ (d\leq D)$, where $^\intercal$ denotes transpose, $(u,v)$ are the horizontal and vertical positions of the top-left point, $(w, h)$ are the width and height, and $D$ is the number of detected faces.
The central point of detection is thus computed as:
\begin{equation}\label{eq:bboxcentral}
    {\boldsymbol{\mu}_{d}}=(u+\frac{1}{2}w, v+\frac{1}{2}h)^\intercal_{d}
\end{equation}

The visual feature is encoded as the exponential part of the multi-variant Gaussian distribution (in $u$ and $v$ direction) with the standard deviations specified by the detection width and height and achieves the maximum at the  central point:
\begin{equation}\label{eq:visualfeature}
\mathcal{V}({\bf x})=
\left\{
\begin{matrix}
\max_{d}
\  e^{ -\frac{1}{2}\left({\bf x}-{\boldsymbol{\mu}_{d}}\right)\Sigma^{-1}_{d}
\left({\bf x}- {\boldsymbol{\mu}_{d}} \right)^\intercal} & D>0, \\ 
\mathcal{U}({\bf x}) & otherwise
\end{matrix}\right.
\end{equation}
where ${\bf x}$ indicates the potential image positions, $\Sigma_{d}=diag(w^2_d, h^2_d)$ is a diagonal covariance matrix, and $\mathcal{U}(\bf x)$ indicates uniform distribution. 
The components in $\mathcal{V}({\bf x})$ are re-sampled to the same length of GCC-PHAT. 


\begin{figure}[!t]\label{tab:video encoding}
\begin{center} 
\subfigure[face detections]{\label{subfig:pic}
\includegraphics[width=0.485\columnwidth]{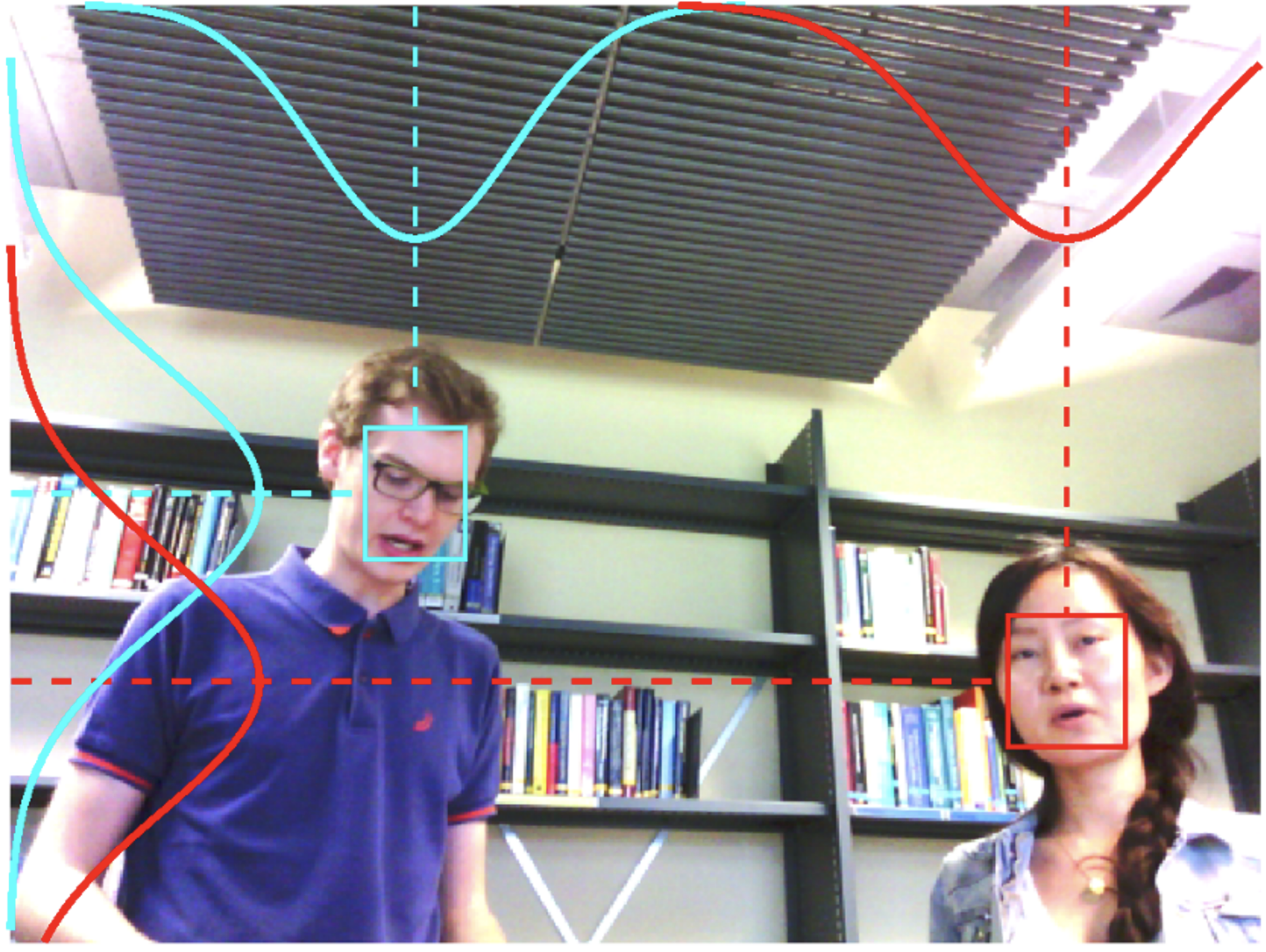}}
\subfigure[visual feature encoding]{\label{subfig:encoding}
\includegraphics[width=0.485\columnwidth]{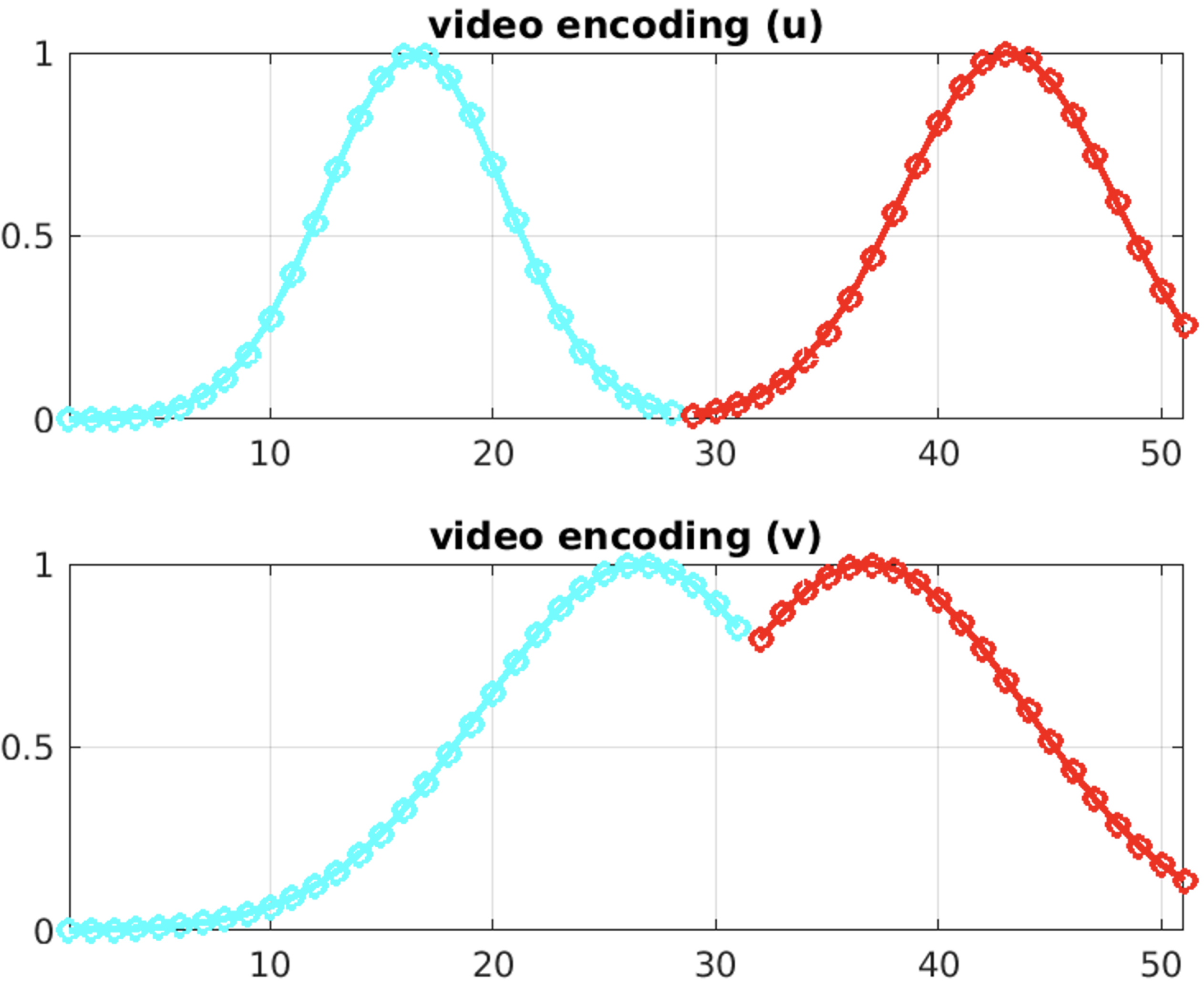}}
\end{center}
  \caption{Visual feature encoding from face detection bounding boxes. The feature resembles the horizontal (top) and vertical (bottom) axis of the image.} 
  \label{fig:NNs}
\end{figure}

\begin{figure}[!htb]
\begin{center} 
\subfigure[]{ \label{subfig:videogeneration}
\includegraphics[width=.95\columnwidth]{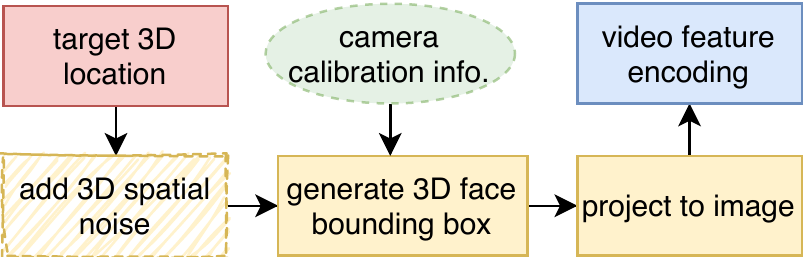}}
\subfigure[]{ \label{subfig:camprojection}
\includegraphics[width=.95\columnwidth]{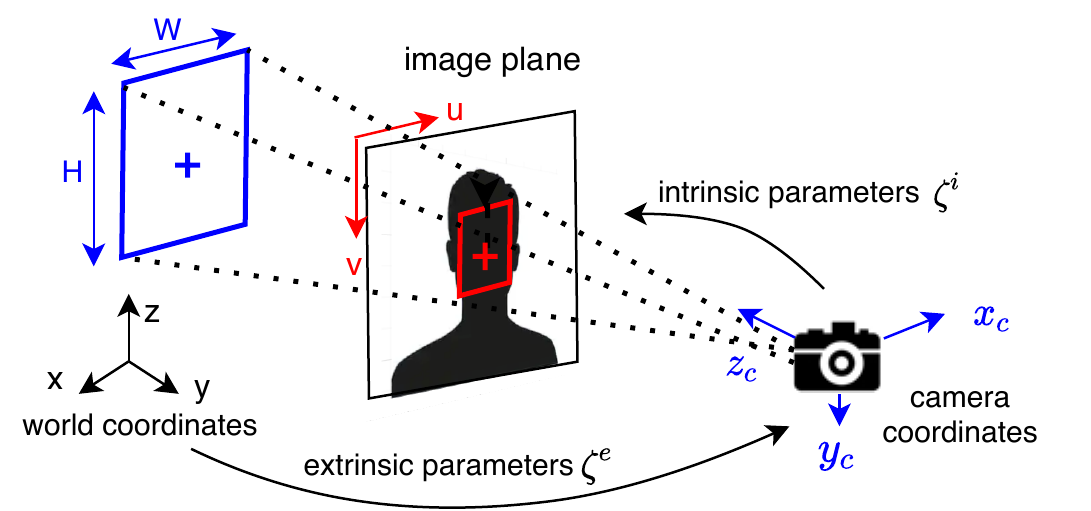}}
\end{center}
  \caption{(a) Pipeline to generate face bounding boxes and visual features and (b) 3D-to-image bounding box projection. $(x,y,z)$: world coordinates; $(x_c, y_c, z_c)$:  camera coordinates; $(u, v)$: image coordinates. }
  \label{fig:cameraprojection}
\end{figure}


Audio-visual parallel data are not abundantly available. However, it is possible to obtain the camera's extrinsic and intrinsic calibration parameters $\mathbf{\zeta}^e$ and $\mathbf{\zeta}^i$, the 3D location ${\bf p}=(x,y,z)^\intercal$ of a sound source. We propose a novel method to synthesize visual features in synchrony with the audio features by \eq~\ref{eq:visualfeature}. The overall pipeline of visual feature generation is illustrated in \fig~\ref{subfig:videogeneration} and the process is formulated next. 

We first add three-variant Gaussian distributed spatial noise
to the target 3D location ${\bf p}$ to account for possible face detection error, and transfer the resulting point to the camera coordinates given the extrinsic parameters:
\begin{equation}\label{eq:3Dtransfer}
        \Tilde{{\bf p}}_c = \Phi (\mathcal{N}({\bf p},\Sigma_{p}) \ | \ \mathbf{\zeta}^e)
\end{equation}
with noise covariance matrix $\Sigma_{p}=diag(\sigma_x^2, \sigma_y^2, \sigma_z^2)$ assuming that the additive noises to $(x,y,z)$ are independent, and $\Phi$ is the transformation using the pin-hole camera model \cite{hartley2003multiple}.

Then, we geometrically create the 3D face bounding box whose plane is perpendicular to the camera's optical axis ($z_c$ in \fig~\ref{subfig:camprojection}), and  project to the image plane:
\begin{equation}\label{eq:imageprojection}
    \chi = \Psi(\Tilde{\bf p}_c + \mathbf{v} \ | \ \mathbf{\zeta}^i)
\end{equation}
where  $\Psi$ is the 3D-to-image projection, $\mathbf{v}$ is the translation vector which equals to $ (-\frac{W}{2},-\frac{H}{2},0)^\intercal$ for the top-left point $\chi^{tl}$ and $(\frac{W}{2},\frac{H}{2},0)^\intercal$  for the bottom-right point $\chi^{br}$, respectively. $W$ and $H$ are the width and height assumptions of a real human face.

Finally, the simulated face detection bounding box ${\bf b}$ is computed as $  {\bf b} = cat(\chi^{tl},  \chi^{br}-\chi^{tl}) $,  where $ cat$ denotes a concatenation operation to form a column vector.

\subsection{Neural network architecture}
We propose two \ac{NN} architectures for audio-visual speaker \ac{DoA} estimation based on \ac{MLP}, namely \ac{MLPAVC} and \ac{MLPAVAW}, which specify different ways of audio-visual feature fusion and classifier design as illustrated in \fig~\ref{fig:NNs}.  

\begin{figure}[!tb]
\begin{center} 
\subfigure[MLP-AVC]{ \label{subfig:MLP-AVC}
\includegraphics[width=0.46\columnwidth]{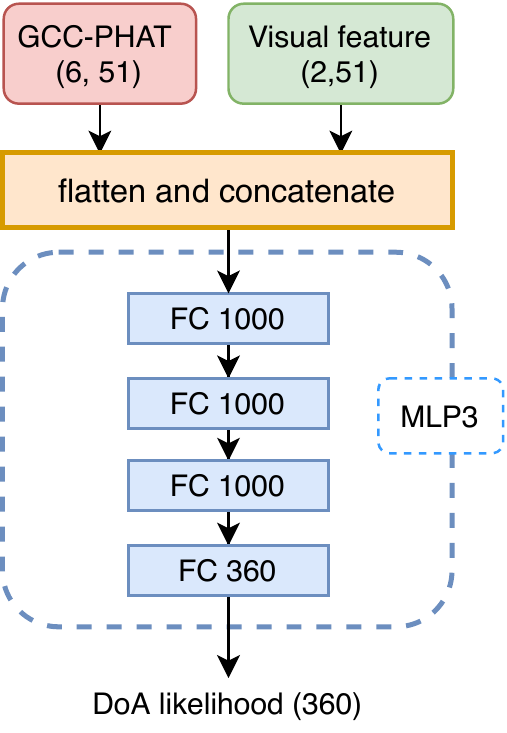}}
\subfigure[MLP-AVAW]{\label{subfig:MLP-AVAW}
\includegraphics[width=0.485\columnwidth]{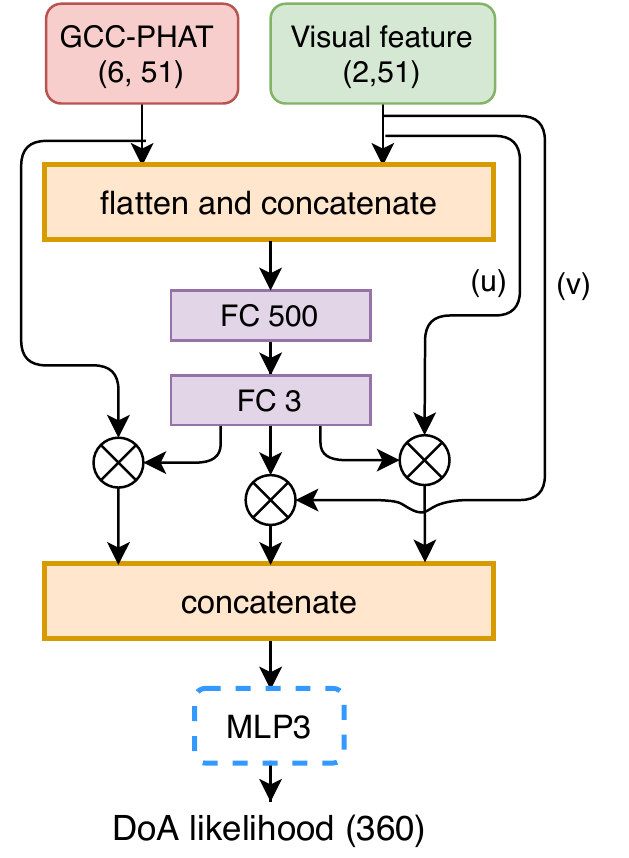} }
\end{center}
  \caption{Proposed \ac{NN} architectures for $360^\circ$ DoA estimation (red: audio block; green: video block; blue: standard \ac{MLP} network; purple: adaptive weighting block; orange: feature reformatting block). The input dimension $(6,51)$ represents 51 GCC-PHAT coefficients for each of the 6 microphone pairs and $(2, 51)$ represents 51 visual feature encoding for the image horizontal and vertical directions.}
  \label{fig:NNs}
\end{figure}

 \ac{MLPAVC} consists of three hidden layers, denoted as MLP3 in \fig~\ref{subfig:MLP-AVC} by a dotted blue box, each one  is a fully-connected layer with ReLU activation \cite{nair2010rectified} and batch normalization~  \cite{ioffe2015batchnorm}. It takes the flattened and concatenated GCC-PHAT and visual features as an input vector. The network is trained to predict the probability of \ac{DoA} labels, as in \cite{he2018deep}, using a sigmoid output layer.  
\ac{MLPAVC} adopts an early fusion strategy by concatenating audio and visual features. We hypothesize that such early fusion doesn't learn to pay selective attention to uni-modal features, that are crucial in face of missing data or noisy data. 

\ac{MLPAVAW} introduces an adaptive weighting mechanism, which uses a tiny NN with two fully-connected layers, colored in purple in \fig~\ref{subfig:MLP-AVAW}), to learn three adaptive weights for the audio GCC-PHAT feature, video image horizontal and vertical features, respectively.
 A softmax activation function is applied for weights normalization. We call this as  `adaptive weighting' mechanism as the weights are adapted according to the live input during inference. Finally, the weighted multi-modal features are  concatenated for MLP3 to compute DoA. 

\begin{table}[!tb]
\centering
\caption{
MAE ($^\circ$) and ACC (\%) of the noisy target 3D locations ($\mathcal{N}({\bf p},\Sigma_{p})$ in \eq~\ref{eq:3Dtransfer}) for visual feature generation of the loudspeaker cases. Results are measured on frames accounting into \ac{DR}.}
\begin{tabular}{|c|c|c|c|c|c|c|}
\hline
\multicolumn{3}{|c|}{\textbf{Train (loudspeaker)}}       & \multicolumn{3}{c|}{\textbf{Test-loudspeaker} }                                             \\ \hline \hline
DR & MAE                  & ACC    &      DR &         MAE                 & ACC                  \\  \hline
11.3 \% & 6.67                & 46.0\%        & 9.2 \%      & 6.28               & 48.9 \%                 \\ \hline
\end{tabular}
\label{tab:facesimulation}
\end{table}

\begin{table*}[!tb]
\centering
\caption{A summary of MAE ($^\circ$) and ACC (\%) of speaker \ac{DoA} estimation on the SSLR test set ($N$ indicates the number of speakers; the number of audio frames for each subset is given in bracket). We reproduce the results of \cite{he2018deep} for comparison.}
\begin{tabular}{cl||c|c|c|c|c|c|c|c||c|c|c|c|}
 \cline{3-12} 
                                        & \multicolumn{1}{c|}{}                                       & \multicolumn{4}{c|}{\textbf{Loudspeaker}}                                                   & \multicolumn{4}{c||}{\textbf{Human}}                                         & \multicolumn{2}{c|}{\textbf{Overall}}                                                      \\ \cline{3-10} 
                                        & \multicolumn{1}{c|}{}                                       & \multicolumn{2}{c|}{\textbf{N=1   (178k)}}                                  & \multicolumn{2}{c|}{\textbf{N=2   (29k)}}                                   & \multicolumn{2}{c|}{\textbf{N=1   (788)}}                                   & \multicolumn{2}{c||}{\textbf{N=2   (141) }}                                   & \multicolumn{2}{c|}{}                                           \\ \cline{3-12} 
     & \multicolumn{1}{c|}{\multirow{-3}{*}{ }} & \multicolumn{1}{l|}{MAE}             & \multicolumn{1}{l|}{ACC}             & \multicolumn{1}{l|}{MAE}             & \multicolumn{1}{l|}{ACC}             & \multicolumn{1}{l|}{MAE}             & \multicolumn{1}{l|}{ACC}             & \multicolumn{1}{l|}{MAE}             & \multicolumn{1}{l||}{ACC}             & \multicolumn{1}{l|}{MAE}             & \multicolumn{1}{l|}{ACC}             \\ \hline \hline
                                 \multicolumn{1}{|l}{}            & \multicolumn{1}{|c||}{SRP-PHAT \cite{brandstein1997robust}    }                                          & {19.00}         & {82.0}          & {36.95}         & {50.0}          & {2.62}          & {93.0}          & {20.90}         & {56.0}            & {21.44}         & {78.0}          \\ \cline{2-12} 
\multicolumn{1}{|c}{\multirow{-2}{*}{\textbf{audio}} }       & \multicolumn{1}{|l||}{MLP-GCC \cite{he2018deep}}                                                   & { 4.06}             & {94.9} & {8.10}          & {71.5}          & {4.75}          & {95.1}          & {5.98}          & {75.5}                 & {4.63}          & 91.6 \\ \hline
          \multicolumn{1}{|l}{}    &    \multicolumn{1}{|l||}{MLP-AVC }                                                  & {3.87}          & {94.8} & {7.80}          & {71.9}          & \textbf{1.84}          & {97.1}          & {3.89} & {81.9}          & {4.42}                  & 91.7 \\ \cline{2-12} 
   \multicolumn{1}{|l}{  \multirow{-2}{*}{\textbf{audio-visual}}} & \multicolumn{1}{|l||}{MLP-AVAW    }                                               & {\textbf{3.73}} & {\textbf{95.0}} & {\textbf{7.28}} & {\textbf{73.6}} & {2.04} & {\textbf{98.0}} & \textbf{3.49}          & {\textbf{86.5}}  & {\textbf{4.22}} & {\textbf{92.0}} \\ \hline
\end{tabular}
\label{tab:results}
\end{table*}

\begin{table*}[!tb]
\centering
\caption{A summary of MAE ($^\circ$) and ACC (\%) of speaker \ac{DoA} estimation on the SSLR test set with different \ac{SNR}s and face detection swap percentage (FDSP)s. The results are obtained using the \ac{MLPAVAW} network architecture.
}
\begin{tabular}{cc|c|c||c|c|c|c|c|c|c|c|c|c|}
\cline{3-14}
\multicolumn{1}{l}{} & \multicolumn{1}{l|}{} & \multicolumn{2}{c||}{\multirow{2}{*}{\textbf{MLP-GCC \cite{he2018deep}}}} & \multicolumn{10}{c|}{\textbf{ Face Detection Swap Percentage }} \\ \cline{5-14}
 &  & \multicolumn{2}{c||}{} & \multicolumn{2}{c|}{\textbf{0\%}} & \multicolumn{2}{c|}{\textbf{10\%}} & \multicolumn{2}{c|}{\textbf{30\%}} & \multicolumn{2}{c|}{\textbf{50\%}} & \multicolumn{2}{c|}{\textbf{70\%}} \\ \cline{3-14}
&  & \textbf{MAE} & \textbf{ACC} & \textbf{MAE} & \textbf{ACC} & \textbf{MAE} & \textbf{ACC} & \textbf{MAE} & \textbf{ACC} & \textbf{MAE} & \textbf{ACC} & \textbf{MAE} & \textbf{ACC} \\ \hline\hline
 \multicolumn{1}{|c|}{\multirow{7}{*}{\textbf{\begin{tabular}[c]{@{}c@{}}audio \\ \\ 
 SNR ($dB$)\end{tabular}}}}  
& \textbf{-10} &   52.74 & 24.2 & 49.69 & 26.59 & 49.98 & 26.5 & 50.51 & 26.2 & 50.87 & 26.0 & 51.21 & 25.8 \\ \cline{2-14}
\multicolumn{1}{|c|}{} & \textbf{0} & 26.19 & 54.8 & 22.19 & 57.3  & 22.37 & 57.2 & 22.73 & 57.0 & 23.01 & 56.7 & 23.22 & 56.6 \\ \cline{2-14}
\multicolumn{1}{|c|}{} & \textbf{10} & 10.64 & 78.8 & 9.34  & 79.2  & 9.41  & 79.1 & 9.52  & 79.0 & 9.61  & 78.9 & 9.67  & 78.8 \\ \cline{2-14}
\multicolumn{1}{|c|}{} & \textbf{20} & 6.02  & 89.0 & 5.68  & 89.0  & 5.70  & 89.0 & 5.75  & 89.0 & 5.79  & 88.9 & 5.82  & 88.9 \\ \cline{2-14}
\multicolumn{1}{|c|}{} & \textbf{Clean} &4.63  & 91.6 & 4.22  & 92.0  & 4.24  & 92.0 & 4.28  & 91.9 & 4.31  & 91.9 & 4.32  & 91.9\\  \hline  
\end{tabular}
\label{tab:results_MAE}
\end{table*}

\section{Experiments}

\subsection{Dataset and performance metrics}
The existing audio-visual datasets, such as AV16.3 \cite{lathoud2004av16}, CAV3D \cite{qian2019multi}, and AVASM \cite{deleforge2015co}, are either of limited size, or don't provide the spatial ground truth. We, therefore, simulate the synchronized visual features for a \ac{SSL} dataset of the loudspeaker cases. We choose the recently released SSLR dataset\footnote{SSLR dataset: \url{https://www.idiap.ch/dataset/sslr/}} 
 \cite{he2018deep}, that is recorded in a physical setup from one or two concurrent speakers, and with adequate target 3D annotations.
It consists of 4-channel audio recordings at $48 \ kHz$ sampling rate, that is organized into three subsets, namely train (loudspeaker), test-human, and test-loudspeaker.

\begin{figure}[!tb]
\begin{center} 
\subfigure[camera and target 3D locations]{\label{subfig:SSLcameradata}
\includegraphics[width=0.9\columnwidth]{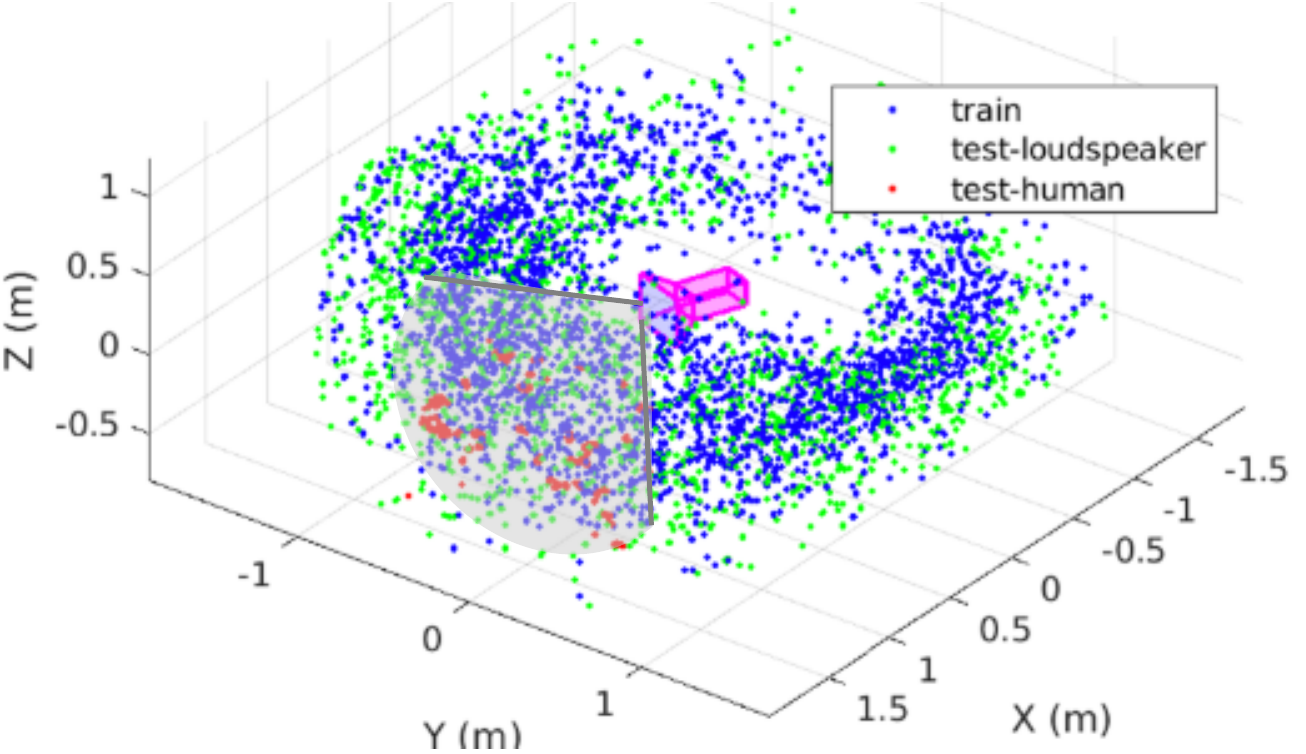}}
\subfigure[train]{\label{subfig:deGCF1}
\includegraphics[width=0.31\columnwidth]{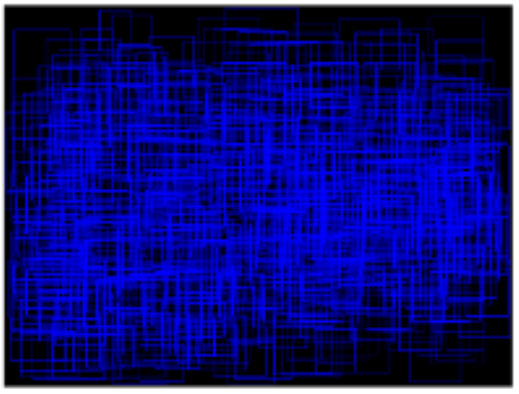}}
\subfigure[test-loudspeaker]{\label{subfig:deGCF2}
\includegraphics[width=0.31\columnwidth]{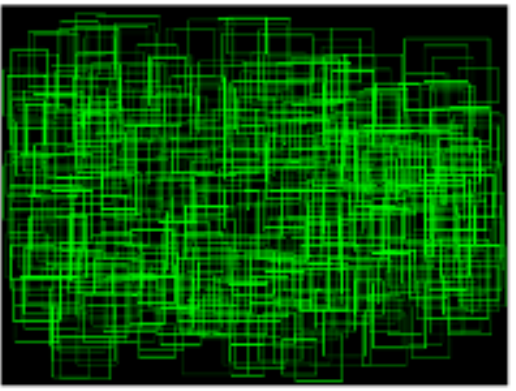}}
\subfigure[test-human]{\label{subfig:deGCF2}
\includegraphics[width=0.31\columnwidth]{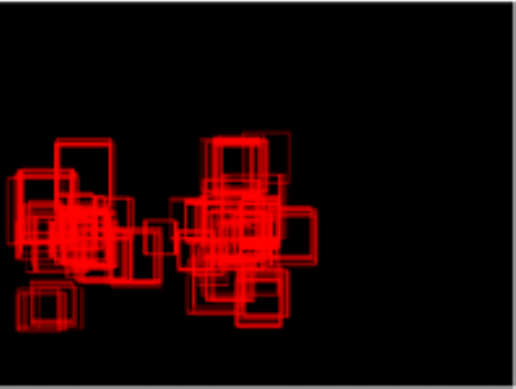}}
\end{center}
  \caption{(a) Camera and target 3D locations (the gray section indicates the camera's \ac{FoV}); (b-c) The distribution of the projected face detection bounding boxes (from points in gray region in (a)) on the image plane of different SSLR subsets (blue: train; green: test-loudspeaker); (d) RetinaFace detections \cite{Deng_2020_CVPR} on test-human.}
  \label{fig:detection}
\end{figure}


 We evaluate the performance of \ac{DoA} estimates using the same metrics of \cite{he2018deep} \ie \ac{MAE} and \ac{ACC}, where \ac{MAE} is defined as the mean absolute error between the actual and the estimated DoA, while the accuracy allowance of \ac{ACC} is  $5^\circ$ in the classification prediction.

For the test-human subset, we apply the RetinaFace detector \cite{Deng_2020_CVPR} to achieve the face bounding boxes.
For the train and test-loudspeaker subsets, the visual features are simulated with the method proposed in \Sec~\ref{ssec:videofeature} with a noise covariance matrix $\Sigma_{p}=diag(0.2,0.2,0.2)$.
\fig~\ref{subfig:SSLcameradata} illustrates the ground truth camera (magenta) and target 3D locations for the train (blue), test-loudspeaker (green) and  test-human (red) subsets for all frames. Targets in the gray region are inside the camera's \ac{FoV}, therefore, visible to the camera. We only generate face bounding boxes of visible targets, as visualized in \fig~\ref{fig:detection}(b-c) and formulated in  \eq~\ref{eq:3Dtransfer}-\ref{eq:imageprojection} with the simulated bounding box ${\bf b}$.
 \fig~\ref{fig:detection} shows that the face bounding boxes spread well across the FoV with a balanced distribution.
 We don't generate bounding boxes for speakers that are outside the FoV. As a result, the visual features for the invisible speakers become missing data 
  ({the normal distribution in \eq~\ref{eq:visualfeature} for visual feature representation) 
 in the audio-visual dataset.}

The statistics of simulated visual features are summarized in \tab~\ref{tab:facesimulation}
where \ac{DR} represents the percentage of video frames having targets inside the FoV. Low DR means a high percentage of missing visual features. 
 We also report in \tab~\ref{tab:facesimulation} the \ac{DoA} \ac{MAE} and \ac{ACC} of the simulated visual features, indicating that the simulated data is of enough difficulty to represent real scenarios.




\subsection{Parameter settings}
 The \ac{GCC-PHAT} is computed for every $170 \ ms$ segments  with delay lags $\tau \in [-25,25]$, resulting in 51 coefficients for each microphone pair as in \cite{he2018deep}. 
 With 6 microphone pairs, each pair contributing  51 GCC-PHAT coefficients, we obtain 306 GCC-PHAT coefficients. For visual features, the human face width and height are assumed to have $W=0.14\ m, H=0.18\ m$, respectively as such in \cite{qian2019multi}. We adjust the size of the horizontal and vertical visual feature encoding to 51 to match that of GCC-PHAT coefficients.
 
We use the Adam optimizer \cite{kingma2014adam}. All models are trained for 10 epochs with a batch size of 256 samples and a learning rate of 0.001. 
Since  multi-speaker localization is not a single-label classification problem,  we use \ac{MSE} instead of cross-entropy as the loss function.

\subsection{Results}

\tab~\ref{tab:results} provides the experimental results on the SSLR test set. Results are separately reported for different subsets and the speaker number (assumed to be known). 
The best result for each column is in the bold font.
We compare the results of MLP-AVC and MLP-AVAW with two audio baseline methods: the traditional \ac{SRP-PHAT} method \cite{brandstein1997robust} and the state-of-the-art MLP-GCC method \cite{he2018deep}.  
As speakers are not always visible, we don't provide the video-only baseline to avoid unfair comparison. Furthermore, \tab~\ref{tab:facesimulation} suggests that it is challenging to expect visual features alone to outperform the audio \ac{DoA} estimation.

\tab~\ref{tab:results} shows that, by both early fusion of audio-visual features. In particular, \ac{MLPAVC} reduces \ac{MAE}  from $4.63^\circ$ (MLP-GCC) to $4.42^\circ$, which confirms the audio-visual fusion benefits. For the test-human subset, speakers are mostly inside the camera's FoV (the red points locate in the gray region in \fig~\ref{subfig:SSLcameradata}) and \ac{DR} of the RetinaFace detector \cite{Deng_2020_CVPR} achieves 100 \%, which is much
higher than \ac{DR} in test-loudspeaker (9.2 \%). Thus, the \ac{MAE} degradation in test-human (from $4.75^\circ$ to $1.84^\circ$ and from $5.98^\circ$ to $3.89^\circ$) is more significant than in test-loudspeaker (from $4.06^\circ$ to $3.87^\circ$ and from $8.10^\circ$ to $7.80^\circ$).
Besides, further improvements are introduced by the adaptive weighting mechanism in \ac{MLPAVAW}, which achieves the best results in most cases with the overall \ac{MAE} at $4.22^\circ$ and ACC at $92.0$\%. 

Next, we further evaluate the noise robustness of the proposed networks. 
For audio, we apply additive white Gaussian noise of \ac{SNR}s varying from $-10$ $dB$ to $20$ $dB$ on the original SSLR audio signals. For video, we randomly swap up to $70\%$ face detections to the other frames to generate false positives and false negatives. 
\tab~\ref{tab:results_MAE} lists the overall \ac{MAE} and \ac{ACC}  of \ac{MLPAVAW} in comparison with those under clean audio condition. 
We also provide the MLP-GCC results in the first two columns indicating the audio-only performance without swapping the face detection.
From the results, we can see that
fusing visual features always brings benefits. Additionally, audio is of more importance than video since with the degradation of \ac{SNR}, both \ac{MAE} and \ac{ACC} are getting worse as \ac{FDSP} increases, the performance degradation is also obvious but not so significant.
Even at \ac{FDSP}=$70 \%$, the proposed network still outperforms the MLP-GCC. The performance gains by \ac{MLPAVAW} suggest that visual features provide additional information in degraded acoustic conditions.

\vspace{-.2cm}
\section{Conclusions}
This paper presented two neural network architectures for multi-speaker DoA estimation using audio-visual signals.
The comprehensive evaluation results confirm the benefits of audio-visual fusion and the adaptive weighting mechanism. 
Besides, we proposed a technique to synthesize visual features from 
geometric information about the sound sources to deal with lack of annotated audio-visual data.  
Future work will include exploring network models that can generalize 
with limited training data.


\bibliographystyle{IEEEbib}

\bibliography{main}




\end{document}

%% file: acro.tex
\newacro{TDOA}[TDoA]{Time Difference of Arrival}
\newacro{DoA}[DoA]{Direction of Arrival}
\newacro{AV-GLMB}[AV-GLMB]{Audio-Visual tracking with the Generalized Labelled Multi-Bernoulli}
\newacro{SoA}[SoA]{state-of-the-art}
\newacro{MAE}[MAE]{\textit{Mean Absolute Error}}
\newacro{ACC}[ACC]{\textit{Accuracy}}
\newacro{KF}[KF]{Kalman Filter}
\newacro{EKF}[EKF]{Extended Kalman Filter}
\newacro{DKF}[DKF]{Decentralized Kalman Filter}
\newacro{SSL}[SSL]{Sound Source Localization}
\newacro{PHD}[PHD]{Probability Hypothesis Density}
\newacro{FoV}[FoV]{Field-of-View}
\newacro{PF}[PF]{Particle Filter}
\newacro{PDF}[PDF]{Probability Density Function}
\newacro{GCC}[GCC]{Generalized Cross Correlation}
\newacro{GCF}[GCF]{Global Coherence Field}
\newacro{GCC-PHAT}[GCC-PHAT]{Generalized Cross Correlation with Phase Transform}
\newacro{STFT}[STFT]{Short-Time-Fourier-Transform}
\newacro{GLMB}[GLMB]{Generalized Labelled Multi-Bernoulli}
\newacro{MOT}[MOT]{Multiple Object Tracking}
\newacro{SOT}[SOT]{Single Object Tracking}
\newacro{RFS}[RFS]{Random Finite Set}
\newacro{SMC}[SMC]{Sequential Monte Carlo}
\newacro{OSPA}[OSPA]{Optimal Sub-Pattern Assignment}
\newacro{NN}[NN]{Neural Networks}
\newacro{VAD}[VAD]{Voice Activity Detection}
\newacro{MLP}[MLP]{Multilayer Perceptron}
\newacro{CNN}[CNN]{Convolutional Neural Network}
\newacro{MSE}[MSE]{Mean Square Error}
\newacro{AVDOAL}[AVDOAL]{Audio-Visual Direction-of-Arrival Localizatoin}
\newacro{MLPAVC}[MLP-AVC]{MLP Audio-Visual Concatenation}
\newacro{MLPAVAW}[MLP-AVAW]{MLP Audio-Visual Adaptive Weighting}
\newacro{SRP-PHAT}[SRP-PHAT]{Steered Response Power PHAse Transform}
\newacro{SNR}[SNR]{Signal-to-Noise Ratio}
\newacro{FDSP}[FDSP]{Face Detection Swap Percentage}
\newacro{DR}[DR]{Detection Rate}
\newacro{ITDs}[ITDs]{Interaural Time Differences}
\newacro{ILDs}[ILDs]{Interaural Level Differences}